\documentclass[11pt,a4paper]{article}

\usepackage{fontspec}
\usepackage[titletoc,title,header]{appendix}

\usepackage{amsmath}
\usepackage{amssymb}
\usepackage{amsthm}
\usepackage{mathtools}
\usepackage{xparse}    

\usepackage[margin=1in]{geometry}

\usepackage{graphicx}
\usepackage{xcolor}

\usepackage{hyperref}
\hypersetup{
  colorlinks=true,
  linkcolor=blue,
  citecolor=blue,
  urlcolor=blue,
  pdfauthor={Yoshitsugu Sekine},
  pdftitle={Constructive Quantum Field Theory and Rigorous Statistical Mechanics via Operator Algebras and Probability Theory --- Guiding Principles and Research Perspectives}
}

\usepackage[numbers]{natbib}
\theoremstyle{plain}

\AtEndEnvironment{thm}{\qed}

\AtEndEnvironment{lem}{\qed}

\AtEndEnvironment{prop}{\qed}

\AtEndEnvironment{cor}{\qed}

\AtEndEnvironment{conj}{\qed}

\AtEndEnvironment{fact}{\qed}

\theoremstyle{definition}

\AtEndEnvironment{defn}{\qed}

\AtEndEnvironment{ex}{\qed}

\theoremstyle{remark}

\AtEndEnvironment{rem}{\qed}

\makeatletter
\providecommand*{\dashv}{\mathrel{\mathpalette\@Dashv\vDash}}
\newcommand*{\@dashv}[2]{\reflectbox{$\m@th#1#2$}}
\makeatother
\NewDocumentCommand{\weaknorm}{O{\dbk} m}{#1{#2}} 

\newcommand{\rbkt}[2]{\left( #1,\,#2 \right)} 
\newcommand{\isomto}{\mathrel{\rightarrowtail\kern-1.9ex\twoheadrightarrow}} 
\newcommand{\dbk}[1]{\left\langle #1 \right\rangle} 
\newcommand{\rbk}[1]{\left( #1 \right)} 
\newcommand{\sqbk}[1]{\left[ #1 \right]} 
\newcommand{\fun}[2]{#1 \rbk{#2}} 

\NewDocumentCommand{\imunit}{O{\mathsf{i}}}{#1} 
\NewDocumentCommand{\placeholder}{O{\bullet}}{#1} 
\NewDocumentCommand{\trace}{O{\operatorname{Tr}}}{#1} 
\newcommand{\eqcsq}[1]{\sqbk{#1}} 
\newcommand{\opchern}[1]{\operatorname{ch}} 
\newcommand{\setSymbolDownLeft}[2]{{\vphantom{#2}}_{#1}{#2}} 
\newcommand{\setSymbolUpLeft}[2]{{\vphantom{#2}}^{#1}{#2}} 
\NewDocumentCommand{\agvariety}{O{\mathcal}}{#1} 
\NewDocumentCommand{\cmdrel}{O{\omega}}{#1} 
\NewDocumentCommand{\dfsp}{O{A}}{#1} 
\NewDocumentCommand{\eqcpointed}{O{\eqcsq} m}{#1{#2}_{\ast}} 
\NewDocumentCommand{\fnheaviside}{O{H}}{#1} 
\NewDocumentCommand{\fthol}{O{\mathcal{O}}}{#1} 
\NewDocumentCommand{\ftmero}{O{\mathcal{M}}}{#1} 
\NewDocumentCommand{\grcentralizer}{O{Z}}{#1} 
\NewDocumentCommand{\grmetform}{O{2} m m}{\grmet[#1] \! \rbkt{#2}{#3}} 
\NewDocumentCommand{\grmet}{O{2}}{\setSymbolDownLeft{#1}{g}} 
\NewDocumentCommand{\grnormalizer}{O{N}}{#1} 
\NewDocumentCommand{\gropasym}{O{A}}{#1} 
\NewDocumentCommand{\gropsym}{O{S}}{#1} 
\NewDocumentCommand{\grpermorderedpair}{O{\mathcal{P}}}{#1} 
\NewDocumentCommand{\grsym}{O{\mathfrak{S}} m}{#1_{#2}} 
\NewDocumentCommand{\gtbase}{O{\mathcal}}{#1} 
\NewDocumentCommand{\gtfilter}{O{\mathcal}}{#1} 
\NewDocumentCommand{\gtfmlclosed}{O{\mathcal}}{#1} 
\NewDocumentCommand{\gtfmlopen}{O{\mathcal}}{#1} 
\NewDocumentCommand{\gtopenball}{O{U}}{#1} 
\NewDocumentCommand{\gtopencover}{O{\mathcal}}{#1} 
\NewDocumentCommand{\gtopennbh}{O{\mathcal}}{#1} 
\NewDocumentCommand{\gtpreopencover}{O{\mathcal}}{#1} 
\NewDocumentCommand{\gtsubbase}{O{\mathcal}}{#1} 
\NewDocumentCommand{\gtvicinity}{O{\mathcal}}{#1} 
\NewDocumentCommand{\lasp}{O{\mathcal}}{#1} 
\NewDocumentCommand{\latprightrbk}{O{\top} m}{\rbk{#2}^{#1}} 
\NewDocumentCommand{\latpright}{O{\top} m}{#2^{#1}} 
\NewDocumentCommand{\latp}{O{t} m}{\setSymbolUpLeft{#1}{#2}} 
\NewDocumentCommand{\lpdistribution}{O{\mu} m}{#2_{\ast,#1}} 
\NewDocumentCommand{\lpmollifier}{O{\rho}}{#1} 
\NewDocumentCommand{\lpofpositive}{O{\chi}}{#1} 
\NewDocumentCommand{\manliederiv}{O{L}}{#1} 
\NewDocumentCommand{\mansmoothnbh}{O{\mathcal}}{#1} 
\NewDocumentCommand{\mblfmldsysgenerated}{O{d} m}{\fun{#1}{#2}} 
\NewDocumentCommand{\mblfmlgenerated}{O{\sigma} m}{\fun{#1}{#2}} 
\NewDocumentCommand{\oacorrfn}{O{\Gamma}}{#1} 
\NewDocumentCommand{\oagnsvector}{O{\Omega}}{#1} 
\NewDocumentCommand{\oaideal}{O{\mathcal}}{#1} 
\NewDocumentCommand{\oanumberoperator}{O{A}}{#1} 
\NewDocumentCommand{\oaposcone}{O{\mathcal{P}}}{#1} 
\NewDocumentCommand{\oapressure}{O{P}}{#1} 
\NewDocumentCommand{\oarepn}{O{\pi}}{#1} 
\NewDocumentCommand{\oaspnormalstate}{O{N}}{#1} 
\NewDocumentCommand{\oasppurestate}{O{P}}{#1} 
\NewDocumentCommand{\oaspstate}{O{E}}{#1} 
\NewDocumentCommand{\oastatevector}{O{\Omega}}{#1} 
\NewDocumentCommand{\oastate}{O{\omega}}{#1} 
\NewDocumentCommand{\opdilation}{O{\delta}}{#1} 
\NewDocumentCommand{\opdmat}{O{\rho}}{#1} 
\NewDocumentCommand{\opfockan}{O{a}}{#1} 
\NewDocumentCommand{\opfockcran}{O{a}}{#1^{\#}} 
\NewDocumentCommand{\opfockcrdagger}{O{a}}{#1^{\dagger}} 
\NewDocumentCommand{\opfockcr}{O{a}}{#1^{\ast}} 
\NewDocumentCommand{\opfocknumber}{O{N}}{#1} 
\NewDocumentCommand{\opfocksegalconj}{O{\pi}}{#1} 
\NewDocumentCommand{\opfocksegal}{O{\phi}}{#1} 
\NewDocumentCommand{\opspecmeas}{O{E}}{#1} 
\NewDocumentCommand{\opspec}{O{} m}{\fun{\sigma_{#1}}{#2}} 
\NewDocumentCommand{\optransl}{O{\tau}}{#1} 
\NewDocumentCommand{\opvarspec}{O{} m}{\operatorname{spec}_{#1} #2} 
\NewDocumentCommand{\physaction}{O{\mathcal{A}}}{#1} 
\NewDocumentCommand{\physcharge}{O{e}}{\mathrm{#1}} 
\NewDocumentCommand{\physcplconst}{O{\mathsf{g}}}{#1} 
\NewDocumentCommand{\physelectrostaticcapasity}{O{\mathrm{Cap}}}{#1} 
\NewDocumentCommand{\physenergy}{O{E}}{#1} 
\NewDocumentCommand{\physgse}{O{E}}{#1_{0}} 
\NewDocumentCommand{\physham}{O{H}}{#1} 
\NewDocumentCommand{\physlagdensity}{O{\mathcal{L}}}{#1} 
\NewDocumentCommand{\physlag}{O{L}}{#1} 
\NewDocumentCommand{\physliouvilean}{O{L}}{#1} 
\NewDocumentCommand{\physmass}{O{m}}{#1} 
\NewDocumentCommand{\prbbrownmv}{O{B}}{#1} 
\NewDocumentCommand{\prbcharfun}{O{\chi}}{#1} 
\NewDocumentCommand{\prbdist}{O{\mathcal{P}}}{#1} 
\NewDocumentCommand{\prbgaussianmeasure}{O{\msrcal{N}}}{#1} 
\NewDocumentCommand{\prbnormaldist}{O{N}}{#1} 
\NewDocumentCommand{\prbpoissonprocess}{O{N}}{#1} 
\NewDocumentCommand{\prbprocess}{O{X}}{#1} 
\NewDocumentCommand{\prbqspace}{O{\mathcal{Q}}}{#1} 
\NewDocumentCommand{\prbspsample}{O{\Omega}}{#1} 
\NewDocumentCommand{\psh}{O{\mathfrak}}{#1} 
\NewDocumentCommand{\qtquantumchannel}{O{\mathcal{L}}}{#1} 
\NewDocumentCommand{\repn}{O{\pi}}{#1} 
\NewDocumentCommand{\schattencls}{O{\mathbb{K}}}{#1} 
\NewDocumentCommand{\setfmlcylinder}{O{\mathcal{C}}}{#1} 
\NewDocumentCommand{\setfml}{O{\mathcal}}{#1} 
\NewDocumentCommand{\setindex}{O{\mathcal} m}{#1{#2}} 
\NewDocumentCommand{\setlattice}{O{\Gamma}}{#1} 
\NewDocumentCommand{\setspecial}{O{\mathcal} m}{#1{#2}} 
\NewDocumentCommand{\shdiffform}{O{\sheaf{A}}}{#1} 
\NewDocumentCommand{\sheaf}{O{\mathfrak}}{#1} 
\NewDocumentCommand{\smchemicalpotential}{O{\mu}}{#1} 
\NewDocumentCommand{\smenergydensity}{O{\varrho}}{#1} 
\NewDocumentCommand{\smfluctuationwithdmat}{O{\beta} m}{\smuncertaintywithdmat[#1]{#2}^2} 
\NewDocumentCommand{\sminvtemperature}{O{\beta}}{#1} 
\NewDocumentCommand{\smlocaldensityoperator}{O{\rho}}{#1} 
\NewDocumentCommand{\smmicrocanonicalstate}{O{\beta} m}{\physmean{#2}_{#1}} 
\NewDocumentCommand{\smnumberdensity}{O{\rho}}{#1} 
\NewDocumentCommand{\smooth}{O{\mathcal{E}}}{#1} 
\NewDocumentCommand{\smparticlenumber}{O{N}}{#1} 
\NewDocumentCommand{\smpressure}{O{p}}{#1} 
\NewDocumentCommand{\smspecificfreeenergy}{O{\bar{f}}}{#1} 
\NewDocumentCommand{\smthermalvac}{O{\beta}}{\Omega_{#1}} 
\NewDocumentCommand{\smuncertaintywithdmat}{O{\beta} m}{\rbk{\triangle #2}_{#1}} 
\NewDocumentCommand{\sphilbfrak}{O{\mathfrak}}{#1} 
\NewDocumentCommand{\sphilb}{O{\mathcal}}{#1} 
\NewDocumentCommand{\splowerhalf}{O{\mathbb{H}}}{#1_{\txtneg}} 
\NewDocumentCommand{\spupperhalf}{O{\mathbb{H}}}{#1_{\txtnonneg}} 
\NewDocumentCommand{\topmetric}{O{d}}{#1} 
\NewDocumentCommand{\vaoutnormal}{O{\widehat}}{#1} 
\newcommand{\category}[1]{\mathop{\mathsf{#1}}} 

\newcommand{\catpresheaf}[1]{\category{PSh}} 
\newcommand{\msrcal}[1]{\mathcal{#1}} 

\newcommand{\oacstar}{C^{\ast}} 
\newcommand{\oawstar}{W^{\ast}} 
\newcommand{\oa}[1]{\mathcal{#1}} 
\newcommand{\opspecint}[1]{\mathcal{E}} 
\newcommand{\physmean}[1]{\dbk{#1}} 
\newcommand{\seq}[2]{\if\relax\detokenize{#1}\relax \rbk{#1} \else \rbk{#1}_{#2} \fi} 
\newcommand{\setisomorphism}[1]{\operatorname{Iso}} 
\newcommand{\txtneg}{\mathrm{-}} 
\newcommand{\txtnonneg}{\mathrm{+}} 
\title{Constructive Quantum Field Theory and Rigorous Statistical Mechanics via Operator Algebras and Probability Theory --- Guiding Principles and Research Perspectives}

\author{%
Yoshitsugu Sekine\\{\small\texttt{4429sekine@gmail.com}}%
}

\date{\today}

\begin{document}

\maketitle

\begin{abstract}
We present a hierarchical viewpoint on the operator-algebraic formulation of quantum systems, in which \(C^{*}\)-algebras are responsible for the universal and intrinsic description, whereas von Neumann algebras provide the detailed account obtained after fixing a state compatible with the dynamics. From this standpoint, for bosonic many-body systems the resolvent algebra, rather than the Weyl algebra, is the natural object; in particular, its nuclearity, trivial center, and rich ideal structure faithfully reflect purely quantum-mechanical structures. Macroscopic variables or sector structures associated with phase transitions are captured as the center appearing in the weak closure of the GNS representation. Moreover, the equivalence between representations of operator algebras and functional integrals allows powerful probabilistic methods to be employed. Taking these as guiding principles, we outline research perspectives on concrete objects in constructive quantum field theory and rigorous statistical mechanics.

\end{abstract}

\setcounter{tocdepth}{3}
\tableofcontents

\section{Introduction}\label{introduction}

The purpose of this paper is to reorganize the conceptual significance of operator-algebraic methods, and to clarify, in concrete physical contexts, the division of roles between \(\oacstar\)-algebras and von Neumann algebras in the description of quantum systems. In particular, taking the description of phase transitions---where the striking differences between them are most visible---as an example, we explain the essentially different descriptive powers of these two objects. We then propose research directions in which this feature can be exploited for the thorough analysis of concrete models in constructive quantum field theory and rigorous statistical mechanics, with a view also toward condensed matter theory and open quantum systems.

In quantum statistical mechanics and algebraic quantum field theory, \(\oacstar\)-algebras are generally regarded as the basic starting point, as algebras of quasi-local observables \cite{RudolfHaag001,BratteliRobinson003,BratteliRobinson004}. Together with the control provided by the norm topology and the general formulation of states and dynamics, they yield a universal description of quantum systems. If one views an \(\oacstar\)-algebra as a noncommutative algebra of continuous functions and quantum fluctuations as the absence of discontinuities, then the norm topology is the natural topology that preserves quantum fluctuations, so that \(\oacstar\)-algebras are purely quantum-mechanical objects. By contrast, a von Neumann algebra is not a universal starting point for arbitrary quantum systems; it is an object that already carries information about a specific system and a specific state. For a concrete quantum system one first fixes a dynamics, takes the GNS representation associated with a state compatible with the dynamics---such as a ground state, an equilibrium state, or a vacuum state---and then the von Neumann algebra arises as its weak closure. When the von Neumann algebra has a nontrivial center, its center represents macroscopic observables or the sector structure of the system. Moreover, if the center of an operator algebra represents classicality, then an \(\oacstar\)-algebra, to which we ascribe purely quantum existence, should not possess a nontrivial center.

Rather than leaving all of this as an abstract framework, one should put it to work in the analysis of concrete systems in the style of constructive quantum field theory and rigorous statistical mechanics. In particular, for bosonic many-body systems we adopt, as our main object, the resolvent algebra introduced by Buchholz and his collaborators \cite{BuchholzGrundling002}. The resolvent algebra was proposed as an \(\oacstar\)-algebraic formulation of the canonical commutation relations alternative to the traditional Weyl algebra. It has the advantages of being built solely from bounded operators, of allowing a wide range of dynamics, and of possessing a rich ideal structure \cite{DetlevBuchholz001}. In particular, owing to the faithfulness of its regular representation, the resolvent algebra itself has only a trivial center, and thus represents purely quantum structures. The significance of this construction is best understood from the viewpoint of designing an \(\oacstar\)-algebra that properly reflects physically meaningful operations.

On the other hand, if one considers states accompanied by a phase transition, such as an equilibrium state corresponding to Bose--Einstein condensation, then a nontrivial center emerges in the von Neumann algebra obtained from its GNS representation \cite{ArakiWoods001,AsaoArai028,YoshitsuguSekine004}. This center corresponds to macroscopic quantities such as the order parameter or the phase of the condensate, and is described through a direct integral decomposition. One sees that macroscopic classical variables do not belong to the \(\oacstar\)-algebra itself, but appear in the von Neumann algebra through the representation associated with a physically selected state.

This state- and representation-theoretic structure is more than a mere conceptual reorganization. In particular, the introduction of probability theory as a powerful tool for estimating correlation functions and expectation values can itself be regarded as a form of representation theory \cite{KleinLandau001,LorincziHiroshimaBetz002,LorincziHiroshimaBetz003,DerezinskiGerard001,YoshitsuguSekine004}. In constructive quantum field theory and rigorous statistical mechanics, functional integrals and probability measures play this role, linking operator-algebraic representations with probabilistic descriptions.

The above discussion is not merely an armchair argument: it has in fact been successfully applied to various physical systems \cite{YoshitsuguSekine001,YoshitsuguSekine002,YoshitsuguSekine004}. On this basis we propose the following guiding principle: namely, \(\oacstar\)-algebras are responsible for the universal and intrinsic description of quantum systems, the GNS representation is determined once one fixes a state compatible with the dynamics---such as a ground state, a thermal equilibrium state, or a vacuum state---and the von Neumann algebra then provides the framework for a detailed description of the concrete states of the system under consideration. Furthermore, since the description by von Neumann algebras is equivalent to one by functional integrals \cite{KleinLandau001,DerezinskiGerard001,YoshitsuguSekine004}, the latter provides powerful analytic tools and new physical intuition. This distinction reflects a hierarchical structure that concerns both the concepts physically describable and the concrete and powerful techniques for evaluating limits.

Our standpoint here is not to give priority to mathematical frameworks over physics. Rather, we emphasize the viewpoint of mathematical physics as an endeavor to choose an appropriate mathematical language for a deeper understanding of physics and to sharpen its precision. Operator algebras and probabilistic methods are powerful tools for this purpose, and it is essential to consciously vary the tools depending on the physical phenomenon one wishes to capture.

From this viewpoint, we adopt a standpoint that emphasizes concrete and constructive \(\oacstar\)-algebraic frameworks---such as the resolvent algebra---as the algebras of physical observables. We aim not to stay within abstract general theory, but rather, through actually constructed models, their associated physical states and GNS representations, and the corresponding probabilistic descriptions, to gain a precise understanding of such phenomena as phase transitions, quantum measurement, and the emergence of macroscopic variables. We position this guiding principle, grounded in the concrete results accumulated in constructive quantum field theory and rigorous statistical mechanics, as a starting point from which to drive research forward.

\section{\texorpdfstring{Hierarchy of Description: \(\oacstar\)-Algebras and von Neumann Algebras}{Hierarchy of Description: \textbackslash oacstar-Algebras and von Neumann Algebras}}\label{hierarchy-of-description-oacstar-algebras-and-von-neumann-algebras}

In the operator-algebraic formulation of quantum systems, the distinction between \(\oacstar\)-algebras and von Neumann algebras has been fundamental since early on. The physically important difference lies not in the choice of topology, but in the choice of the object to be discussed and in the stage of description. Since the distinction is easier to clarify in this setting, we take Bose--Einstein condensation---which can be treated in a mathematically rigorous way---as an example when discussing phase transitions in general, although conceptually ferromagnetism has the same structure. A more concrete discussion using the resolvent algebra will be given later.

Let us reconsider the difference between the descriptive power of \(\oacstar\)-algebras for quantum phenomena and the state-dependent descriptive power of von Neumann algebras. As already discussed, an \(\oacstar\)-algebra provides an abstract algebra of observables with purely quantum-mechanical features. A von Neumann algebra is obtained only after one chooses an appropriate representation of the \(\oacstar\)-algebra and takes the weak closure of its image. In physical applications, one typically first specifies dynamics on the \(\oacstar\)-algebra. One then selects a ground state, an equilibrium state, or a vacuum state compatible with those dynamics and forms the corresponding GNS representation. The resulting von Neumann algebra carries information about both the dynamics and the selected state. This dependence makes von Neumann algebras indispensable for detailed studies of concrete systems, but it can make them less suitable for a universal formulation of quantum systems because they retain more dynamical and state-dependent information than such a formulation requires.

The state dependence of a von Neumann algebra can become visible through its center. For example, in the presence of a phase transition, the center may be nontrivial and may encode macroscopically observable classical information. If the universal algebra of observables is intended to retain only quantum-mechanical information rather than classical data associated with a particular state, it is natural to select \(\oacstar\)-algebras with trivial center as the universal starting point. The center of a von Neumann algebra may then carry the macroscopic classical information that emerges only after a state has been selected. The distinction lies in what one takes as the object of discussion, and the difference in topology is a technical device to extract the required information as thoroughly as possible.

Let us first give an example in which the structure of the \(\oacstar\)-algebra itself carries important information. Even without specifying a state or representation, the properties of an \(\oacstar\)-algebra alone can impose constraints on the properties of any system in which it appears; here we discuss the significance of nuclearity of \(\oacstar\)-algebras \cite{MasamichiTakesaki006}. In describing composite systems, one needs to consider tensor products of \(\oacstar\)-algebras. In general, however, the tensor-product norm on \(\oacstar\)-algebras is not uniquely determined, so that fixing the algebraic starting point of a composite system is, mathematically, highly nontrivial. This is where nuclearity enters. Nuclearity removes the ambiguity of the tensor product and serves as a basic condition for formulating composite systems in a natural way. In particular, in connection with the discussion of the relative phase of two Bose--Einstein-condensed systems, we explain how nuclearity determines the macroscopic-variable structure of the composite system.

It is well known that Bose--Einstein condensation occurs in the free Bose gas; the condensed state admits a direct integral decomposition, and the corresponding von Neumann algebra possesses a nontrivial center \cite{ArakiWoods001,BratteliRobinson004,AsaoArai028,YoshitsuguSekine004}. This center represents variables describing the state of condensation, and, in analogy with the magnetization of a ferromagnet, we call it here a macroscopic variable representing the phase transition. The situation is the same for ferromagnetism. In a state with symmetry, components corresponding to various directions of magnetization coexist, and the macroscopically observed magnetization is understood as a decomposition of the state. The center and the direct integral decomposition also play the role of macroscopic variables distinguishing different phases.

The relative phase of two interacting Bose--Einstein-condensed systems has been experimentally measured \cite{HallMatthewsCornellWieman001}. Since the relative phase is also expected, on physical grounds, to be a macroscopic variable, we investigate whether a framework actually yielding such a conclusion can be obtained. Suppose that the \(\oacstar\)-algebras \(\oa{A}_1\) and \(\oa{A}_2\) describing the two systems are both nuclear and have trivial center, that is, are purely quantum-mechanical. The interacting system is then described by the tensor product \(\oa{A}_1 \otimes \oa{A}_2\). Since both factors are nuclear, the tensor product is also nuclear, and its center is trivial. If the state of the system under consideration is faithful, then the GNS representation of the tensor product \(\oacstar\)-algebra is also faithful and has trivial center, so that the relative phase cannot be a quantum observable. Being independent of the details of the system, the possible observables are determined solely by the structure of the \(\oacstar\)-algebra. This is precisely the behavior one should expect of a mathematical object representing physics. Of course, nuclearity is not the only mathematical structure that guarantees such a behavior; nevertheless, since nuclearity also plays an important role in relativistic algebraic quantum field theory \cite{RudolfHaag001}, there are ample reasons to focus on it. The resolvent algebra already mentioned is a nuclear \(\oacstar\)-algebra with trivial center \cite{BuchholzGrundling002,DetlevBuchholz001}, which is yet another reason to pay attention to it.

On the other hand, nuclearity has nothing to do with the emergence of macroscopic variables or centers. While nuclearity stabilizes the purely quantum-mechanical starting point of composite systems, macroscopic quantities such as the relative phase appear only through the choice of a physical state, its GNS representation, and the weak closure. This structure is essential to the understanding of phase transitions and order parameters. Macroscopic variables are not already present inside the algebra; the center emerges, as a property of the system under consideration, in the weak closure of the GNS representation of a state compatible with the dynamics. This is the positive reason for choosing \(\oacstar\)-algebras as the universal starting point and von Neumann algebras in situations in which structural decomposition---as in measurement or phase transitions---is essential. From the discussion above, we obtain the following division of roles: \(\oacstar\)-algebras are responsible for the universal description, and von Neumann algebras for the state-dependent decomposition description.

At this point, one encounters a well-known difficulty. If one starts from a \(\oacstar\)-algebraic description of a system, it is natural to formulate its dynamics as a one-parameter group of automorphisms of the \(\oacstar\)-algebra. However, for the Weyl algebra, which is commonly used to describe bosonic fields, there is a serious limitation from this perspective.

Roughly speaking, the issue concerns the size and structure of the automorphism group of the Weyl algebra. It is not that automorphisms are scarce in a set-theoretic sense; on the contrary, outer automorphisms arising from symplectic transformations and translations are abundant. The difficulty lies in the fact that physically natural Hamiltonian time evolutions are not easily realized, at the \(\oacstar\)-algebraic level, as well-behaved dynamics---particularly as inner automorphisms or norm-continuous one-parameter automorphism groups.

This situation originates from the structural feature that, in the Weyl algebra, the field operators themselves do not belong to the algebra, and the corresponding generators appear only outside it. As a result, physically relevant dynamics are typically given as outer automorphisms, and at the level of the \(\oacstar\)-algebra they do not possess sufficiently regular continuity properties nor a direct description in terms of generators. On the other hand, once a state is fixed and one passes to the associated GNS representation, these dynamics are implemented by unitary operators and appear as natural time evolutions on the Hilbert space.

In this sense, the issue is not merely the size of the automorphism group, but rather which kinds of dynamics can be adequately described at the \(\oacstar\)-algebraic level. From this viewpoint, the Weyl algebra has an intrinsic limitation in that it does not sufficiently accommodate many Hamiltonian dynamics that are physically admissible.

The resolvent algebra \cite{BuchholzGrundling002} was introduced precisely to overcome this difficulty. In contrast to the Weyl algebra, many physically relevant dynamics can be realized as automorphisms that preserve the algebra. At the same time, the resulting time evolution is in general not norm-continuous on the entire algebra; instead, continuity is recovered either on suitable sub-\(\oacstar\)-algebras or at the level of representations associated with physically relevant states. This structure more clearly reflects the hierarchical nature of the operator-algebraic description, in which different levels serve distinct roles.

\section{Construction of the Resolvent Algebra and Its Significance}\label{construction-of-the-resolvent-algebra-and-its-significance}

Not only from the viewpoint of reproducing well-known experimental systems mathematically, but also in view of its importance in relativistic quantum field theory, nuclearity can be regarded as an important requirement on the \(\oacstar\)-algebra of observables. From the viewpoint of enlarging the class of admissible dynamics, the inadequacy of the Weyl algebra is beyond remedy. This is where the resolvent algebra \cite{BuchholzGrundling002} came in. It was originally introduced to describe supersymmetry, but its proponent, Buchholz, has vigorously pursued its applications to quantum statistical mechanics and condensed matter theory \cite{BuchholzGrundling001,DetlevBuchholz001,DetlevBuchholz002,BuchholtzYngvason001}. The resolvent algebra is not merely a technical improvement; it clearly reflects a design philosophy concerning which structures are to be retained at the \(\oacstar\)-algebraic level and which are to be entrusted to states compatible with the dynamics and their GNS representations.

Of course, the Weyl algebra is still widely used as an \(\oacstar\)-algebraic formulation of the canonical commutation relations. The Weyl algebra is constructed from exponentiated bounded operators, and its algebraic definition and manipulation are straightforward. In the representation theory, particularly in regular representations, the field operators are recovered as self-adjoint operators \cite{BratteliRobinson004}. It is important that, once a regular representation is fixed, or in the GNS representation of a regular state, the Weyl algebra and the resolvent algebra carry essentially the same information. Indeed, in a regular representation the field operators are explicitly defined, and through their spectral decomposition there is a mutually recoverable relation between Weyl operators and resolvents. Consequently, at the level of von Neumann algebras the two descriptions agree. In particular, at finite temperature one argues under an equilibrium state by passing to the Araki--Woods representation and its weak closure. Since there are few occasions on which one argues at the \(\oacstar\)-algebraic level, the drawbacks of the Weyl algebra are easily overlooked.

However, this agreement is a phenomenon that occurs only after fixing a state and a representation; the situation changes at the level of the \(\oacstar\)-algebra itself. Before one chooses a state, the Weyl algebra and the resolvent algebra are entirely different algebras, and the difference is not merely a matter of the choice of generators. In particular, the resolvent algebra is designed so as to be compatible with the implementation of the dynamics \cite{BuchholzGrundling002}, and to organize the ideal structure and the class of admissible states and representations \cite{BuchholzGrundling001}, thereby carrying convenient and refined structures for the description of physics at the \(\oacstar\)-algebraic level. In fact, in \cite{YoshitsuguSekine004} the singular subspace associated with Bose--Einstein condensation in the free Bose gas is characterized as an ideal of the resolvent algebra. In a paper currently in preparation, the singular subspace associated with the treatment of the infrared divergence of the van Hove model \cite{AsaoArai026} is also characterized as an ideal of the resolvent algebra.

Moreover, for the treatment of infrared divergences, the resolvent algebra has an advantage over the Weyl algebra. In an infrared divergence the expectation values of field operators diverge. In the Weyl algebra, since field operators are exponentiated, they oscillate and their expectation values are not well defined, whereas in the resolvent algebra field operators appear in the denominator, so that the corresponding expectation values may converge to zero. This is also a strength of description at the \(\oacstar\)-algebraic level.

Beyond nuclearity, the resolvent algebra also possesses the other properties that we discussed in the previous section as desirable for an \(\oacstar\)-algebra of observables. For example, the resolvent algebra is faithfully realized in its regular representation, and its center is trivial. In this sense, at the \(\oacstar\)-algebraic level the resolvent algebra has a purely quantum-mechanical structure, and contains no macroscopic decomposition nor classical structure. To repeat, while nuclearity stably provides the formulation of composite systems, it does not automatically introduce macroscopic structures---a point of the utmost importance.

\section{Representation Theory and Probability Theory: Introducing Concrete and Powerful Techniques for Evaluating Limits}\label{representation-theory-and-probability-theory-introducing-concrete-and-powerful-techniques-for-evaluating-limits}

The passage from an \(\oacstar\)-algebra to a von Neumann algebra is natural when one wishes to study a system in detail. While this framework provides structure, it is insufficient for detailed analysis of concrete models. Here we reinterpret representation theory not as a mere abstract construction, but as a framework of concrete constructions and limit evaluations derived from states compatible with the dynamics, and in that framework we clarify the role played by functional integrals.

For a general \(\oacstar\)-algebra, any state yields a representation via the GNS construction. In this sense, the existence of representations is always abstractly guaranteed. However, this generality is at the same time a limitation. A representation obtained by the GNS construction is structurally correct, but is not necessarily given in a form suitable for concrete analysis or computation. The problem is therefore not whether a representation exists, but which state to choose, which GNS representation to construct, and how to handle it.

The notions of a reference state and a reference representation give physical content to this choice. A paradigmatic example is the Doplicher--Haag--Roberts and Doplicher--Roberts theory in algebraic quantum field theory \cite{DoplicherHaagRoberts003,DoplicherHaagRoberts004,DoplicherRoberts001}. In its standard massive setting, the theory starts from the vacuum representation of the algebra of local observables and describes charged sectors as localized excitations relative to the vacuum sector. The reference representation encodes a concrete physical picture rather than serving as an arbitrary background. When such a picture of a reference state and its excitations is available, the corresponding von Neumann algebras naturally enter a detailed study of the system.

The normal states on a von Neumann algebra in a fixed representation form the folium of that reference representation. A state whose GNS representation is disjoint from the reference representation does not occur as a normal state on this von Neumann algebra. This restriction makes the framework suited to comparing the reference state with states in the same folium, but it excludes states outside that folium, in particular those whose GNS representations are disjoint from the reference representation. A standard infinite-system manifestation is the disjointness of KMS states at different temperatures. More precisely, KMS states for the same dynamics at different temperatures are disjoint when one of them is primary and its GNS von Neumann algebra is of type III \cite[p. 242]{KlausHepp001}. This formulation concerns infinite systems with type III representations rather than finite-dimensional systems.

The abstract \(\oacstar\)-algebra prior to the choice of a representation is not restricted to a single folium. It admits every algebraic state, including states that are not physically selected for a given experimental or thermodynamic situation. In statistical mechanics, typicality gives one mechanism for identifying physically representative behavior and for treating atypical states as negligible \cite{PopescuShortWinter001}. The presence of such unselected states at the abstract level is a feature rather than a defect. It allows one \(\oacstar\)-dynamical system to accommodate mutually disjoint KMS folia at different temperatures without selecting one temperature in advance. A von Neumann algebra associated with a chosen KMS state can then provide the detailed description within that folium.

Von Neumann algebras also interface naturally with probabilistic formulations and perturbation theory. Once a reference state and representation have been fixed, perturbed dynamics and perturbed states can be compared within a common weakly closed algebra. In particular, the perturbation theory of \(\oawstar\)-dynamics provides controlled constructions of perturbed Liouvilleans and KMS states \cite{DerezinskiJaksicPillet001}. This analytic strength accords with the physical picture of a reference state and excitations built upon it.

At this point, probability theory and functional integral theory enter as powerful analytic tools \cite{LorincziHiroshimaBetz002,LorincziHiroshimaBetz003}. As a finite-temperature analogue of the axioms of quantum field theory, a direct correspondence between operator algebras and stochastic processes was already discussed in \cite{KleinLandau001}; however, because generality was emphasized, it was not well suited for the analysis of concrete systems. This situation was substantially improved in the textbook \cite{DerezinskiGerard001}, in which, in particular, the equivalence between the Araki--Woods representation and the Gaussian \(\sminvtemperature\)-Markov path space is clearly formulated. In addition, in \cite{YoshitsuguSekine004}, the operator algebra represented as a direct integral of the Araki--Woods representation, together with the corresponding functional integral---not discussed in the textbook above---is also formulated. In particular, it can handle the case in which the one-particle Hamiltonian is singular and may give rise to infrared divergences, and we are now at the stage where, in a paper currently in preparation, concrete results for interacting models can be presented.

In the treatment of infrared divergences, Bogoliubov and Gross transformations play an important role as transformations linking these representations. They are not merely formal manipulations: they can be used to reorganize the degrees of freedom of the system and to construct effective theories. In this way, not merely representation theory within operator algebras, but its extension combined with probability theory, opens the way to an exhaustive analysis of concrete examples.

The functional integral method provides yet another important viewpoint. In particular, the framework of statistical mechanics---namely, concepts such as Gibbs measures, partition functions, and phase transitions---appears within quantum field theory. Through this correspondence, the theory is not only rigorous, but one also gains an understanding consistent with physical intuition. In other words, operator algebras specify the structure of the system and the framework of observables, representation theory gives their concrete realizations, and functional integrals are responsible for their construction and for the evaluation of limits, thereby providing a full analytic framework. In particular, functional integrals are not an external auxiliary means, but an essential constituent for making the operator-algebraic description actually work. A representation is not merely a realization on a Hilbert space, but the totality of a physical description encompassing a concrete construction supported by the selection of a state compatible with the dynamics, suitable transformations, probability measures, and techniques for evaluating limits. In this sense, abstract operator algebras, concrete representations, and probabilistic methods are mutually complementary.

\section{Guiding Principle: The Hierarchy of Operator-Algebraic Description and Its Extension}\label{guiding-principle-the-hierarchy-of-operator-algebraic-description-and-its-extension}

Let us summarize the hierarchical picture so far. The operator-algebraic description of quantum systems takes the \(\oacstar\)-algebra as its starting point and is made concrete by the choice of a state and the corresponding GNS representation. The von Neumann algebra strongly reflects the properties of the system; in situations where a phase transition occurs, a nontrivial center emerges and macroscopic structures become visible. Nuclearity of the \(\oacstar\)-algebra stabilizes the algebraic starting point of composite systems and provides a foothold for the discussion of composite and interacting systems. Representation theory is not confined to realizations on a Hilbert space, but leads to powerful tools provided by probability measures and functional integrals.

It is worth noting that this hierarchical structure of description is not limited to the emergence of phase transitions or centers, but is already reflected in the very formulation of time evolution itself. In physically relevant situations, dynamics are typically given as outer automorphisms at the level of \(\oacstar\)-algebras, and their full realization only becomes concrete after selecting a state and passing to the corresponding GNS representation.

From this viewpoint, the resolvent algebra may be regarded as a framework designed to make such externally given dynamics accessible already at the \(\oacstar\)-algebraic level. In this sense, the distinction between \(\oacstar\)-algebras and von Neumann algebras should not be understood as merely formal, but rather as corresponding to an intrinsic structural feature inherent in the description of dynamics.

\section{Future Directions}\label{future-directions}

Since several textbooks \cite{AsaoArai026,AsaoArai028,LorincziHiroshimaBetz002,LorincziHiroshimaBetz003,DerezinskiGerard001} have already been published, it is clear that a certain amount of research has accumulated. Even rewriting existing results obtained for bosonic fields using the Weyl algebra in terms of the resolvent algebra, and then thoroughly investigating its ideal structure, would already constitute at least a minimal research program. Furthermore, since models such as the spin--boson model, which describe a rich variety of physics cutting across condensed matter, nonequilibrium statistical mechanics, and open quantum systems, are already being studied within the framework of constructive quantum field theory and rigorous statistical mechanics, the ideal is to develop this line of research by deepening them thoroughly. We now present several problems that the author currently has in mind. The individual problems may overlap.

\subsection{Immediate Problems}\label{immediate-problems}

The preprint \cite{YoshitsuguSekine004} originated from the author's master thesis and its simple finite-temperature extension \cite{YoshitsuguSekine001,YoshitsuguSekine002}. It treats a concrete example of the generalized spin--boson (GSB) model proposed in the papers of Arai and Hirokawa \cite{AraiHirokawa001,AraiHirokawa002}. As shown in \cite{VIYukalov001}, quasi-particles (in particular, acoustic phonons) do not undergo BEC, whereas, as shown in the preprint \cite{YoshitsuguSekine002}, mathematically BEC does occur. In order to investigate this situation, the preprint \cite{YoshitsuguSekine004} deepens the discussion of the van Hove model \cite{AsaoArai026}, which is mathematically almost equivalent, and further descends to the free Bose gas in order to investigate the features of BEC at finite temperature.

Accordingly, the discussion of the van Hove model is the next problem. The functional integral treatment at zero temperature is given in the textbook \cite{LorincziHiroshimaBetz003}, but only as a simple case of the spin--boson model, with no explicit functional-integral discussion specialized to the van Hove model. Although there is no novelty in itself, bearing in mind its role as an introduction to functional integrals---a subject with which the author himself struggled during his own study---we will thoroughly discuss, both at zero temperature and at finite temperature, the treatment via the resolvent algebra and functional integrals.

The essential features of the behavior of the bosonic field in the Hubbard--phonon interacting system \cite{YoshitsuguSekine004} completely coincide with those of the van Hove model. Although restricted to either the Hubbard--phonon interacting system or the van Hove model, we are ready to discuss a no-go theorem for phonon BEC at finite temperature.

The discussion of the spin--boson model, an important model already mentioned, is also indispensable. The theory of functional integrals at zero temperature is discussed in a full chapter of the textbook \cite{LorincziHiroshimaBetz003}. Operator-algebraic results at finite temperature are discussed in the paper \cite{FannesNachtergaeleVerbeure001}. In the general theory of operator algebras, ground states are captured as the \(\sminvtemperature
\to \infty\) limit of \(\sminvtemperature\)-KMS states, so from this viewpoint it is also important to unify the discussion of ground states and equilibrium states of the spin--boson model. Of course, we also wish to extend the no-go theorem for phonon BEC, established for the van Hove model, to the spin--boson model.

Another topic is the discussion of the Luttinger liquid. By the boson--fermion correspondence \cite{AcerbiMorchioStrocchi001,AcerbiMorchioStrocchi002,BuchholzMackTodorov001}, the Luttinger liquid can be described by the resolvent algebra. Moreover, since the Luttinger liquid is also related to conformal field theory, one may be able to make use of the accumulated results on conformal field theory in algebraic quantum field theory. The operator-theoretic paper \cite{LangmannMoosavi001} discusses a fermion--phonon interacting system, which also aligns with the author's original interest in the physical properties of electrons. How---or whether---the physical properties of the Luttinger liquid are inscribed in the ideal structure of the resolvent algebra is also a touchstone for exploring the possibilities of the resolvent algebra.

\subsection{Constructive Quantum Field Theory as a Sobolev Representation Theory}\label{constructive-quantum-field-theory-as-a-sobolev-representation-theory}

In the general theory of operator algebras, the \(\sminvtemperature
\to \infty\) limit of a \(\sminvtemperature\)-KMS state is a ground state. In this sense, one can consider a certain parallel correspondence between ground states and equilibrium states. On the other hand, there is an extremely powerful perturbation theory for equilibrium states, such as the one in the paper \cite{DerezinskiJaksicPillet001}, so one can envisage a route that first discusses equilibrium states and then studies the ground state obtained as a limit. In practice, there are various difficulties in the discussion at the equilibrium-state level and this is not easy, but the operator-algebra/functional-integral correspondence in the textbook \cite{DerezinskiGerard001} is a powerful prescription for overcoming these difficulties.

To develop this argument further, we add another viewpoint, which the author tentatively calls Sobolev representation theory. In ordinary physical quantum mechanics, Dirac's delta function plays a certain role, and the theory of distributions was born out of this. Furthermore, in pursuing the theory of the Schrödinger equation as a partial differential equation, the theory of Sobolev spaces was born as a more tractable distribution theory. The theory of Sobolev spaces has developed explosively, and has itself become an enormous field. We would like to consider its quantum-field-theoretic analogue.

Note that Dirac's delta function is a functional (distribution). A typical example of operator-algebraic infrared-divergence treatment is as follows: one constructs, on Fock space, the ground-state vector of a regularized model, from this ground-state vector one constructs a vector state of the operator algebra acting on Fock space, and then one discusses the ground state of the original model by taking the limit in which the regularization is removed. In this sense, the operator algebra, and in particular its universal representation, can be regarded as the space of distributions in quantum field theory. Of course, like the space of distributions, this space is extremely huge and hard to handle. It is natural to ask whether one can extract from it a more tractable space, analogous to Sobolev spaces. For instance, one can consider auxiliary variables playing the role of the Sobolev exponent, such as temperature. As discussed in the main text, in practice one also needs to consider a suitable representation, and it is this situation that we call Sobolev representation theory.

As promising candidates for Sobolev representation theory, there are already the Araki--Woods representation and functional integral representations. As can be seen from the textbook \cite{LorincziHiroshimaBetz003}, for models in which the interaction is linear in the field---such as the van Hove model, the spin--boson model, and the Nelson model---there is a common prescription for handling infrared divergences, whereas the Pauli--Fierz model has a different character in that it has no infrared divergence. We would like to classify these characteristics appropriately on the basis of concrete models and to exploit the classification in the analysis of other models. For the Pauli--Fierz model, we would like to consider a model obtained by formally adding to the van Hove model a term quadratic in the field, and examine its similarity to the infrared behavior of the Pauli--Fierz model.

It is not clear whether it can be directly connected, but one can also envisage problems such as the identification of spectral subspaces for one-electron states of a relativistic electron in the context of the stability of matter \cite[Chapter 10]{LiebSeiringer001}. By Tomita--Takesaki theory, the spectrum of the Liouvillian at finite temperature is symmetric about the origin. On the other hand, the spectrum of the Dirac equation for a relativistic electron is also expected to be symmetric about the origin. Although it is currently an open problem, there are discussions of the vacuum state in relativistic quantum field theory, and the discussion of spectral subspaces of relativistic electrons is expected not to be unrelated to it. Setting aside the difficulty, from the representation-theoretic viewpoint, representation theory for relativistic quantum field theory is naturally within the scope.

\subsection{Unified Theory of Ground States, Equilibrium States, and Weights}\label{unified-theory-of-ground-states-equilibrium-states-and-weights}

In the doctoral thesis \cite{MartinPorrmann001} on relativistic algebraic quantum field theory, scattering theory is discussed, and there weights are introduced as infraparticle weights. At present, nonrelativistic constructive quantum field theory and rigorous statistical mechanics proceed in terms of states, but when one considers the treatment of infrared divergences, it may be more appropriate to discuss suitable weights.

Although this is not discussed very explicitly in the textbook \cite{AsaoArai026}, if one reformulates it operator-algebraically, then under infrared singular conditions the algebra of observables is contracted in order to exclude states exhibiting infrared divergences, and one then considers states on this contracted algebra of observables. Whether anything new emerges if one instead regards them as weights on the un-contracted algebra of observables from the beginning is, also from the viewpoint of Sobolev representation theory, worth examining. There is also the fact that Tomita--Takesaki theory has been generalized to weights. One possible direction is to extend \cite{KleinLandau001,DerezinskiGerard001} to weights.

There are also problems related to the degeneration of representations, which are also connected with Sobolev representation theory. As is evident from the definition, in the zero-temperature limit the Araki--Woods representation reduces to the Fock representation. As shown in the paper \cite{AsaoArai029}, the Pauli--Fierz model is obtained as the nonrelativistic limit of the relativistic Dirac--Maxwell equation, and here too the relativistic four-component structure degenerates into a two-component structure in the nonrelativistic limit. Here, again, a similarity between the zero-temperature limit and the nonrelativistic limit is observed.

\subsection{\texorpdfstring{Ideal Structure and Representation Theory of Resolvent Algebras and \(\oacstar\)-Algebras}{Ideal Structure and Representation Theory of Resolvent Algebras and \textbackslash oacstar-Algebras}}\label{ideal-structure-and-representation-theory-of-resolvent-algebras-and-oacstar-algebras}

A typical aim is to rewrite, for bosonic fields, existing results obtained via the Weyl algebra in terms of the resolvent algebra, and to study its ideal structure. In particular, as an aspect of the ideal structure of the resolvent algebra under infrared singularity, there is the problem of treating BEC and infrared divergence simultaneously, which is precisely the problem mentioned as the next research topic.

Furthermore, we are interested in the stability of the ideal structure of the resolvent algebra under perturbations. We are interested in when perturbative stability or instability can occur, and in differences between the zero-temperature and finite-temperature cases. In the no-go theorem for quasi-particle BEC currently in preparation, one observes a structure in which infrared singularity kills the singularity of BEC.

Furthermore, besides the resolvent algebra, one can explore whether similar structures appear in CAR algebras or spin systems, or whether there exist other suitable algebras.

\subsection{Reformulation of Operator-Algebraic Structures via Probability Theory and Functional Integrals}\label{reformulation-of-operator-algebraic-structures-via-probability-theory-and-functional-integrals}

This is a topic with a relatively strong mathematical flavor.

In \cite{KleinLandau001,DerezinskiGerard001}, although restricted to the operator algebra corresponding to bosonic fields, there is a discussion of Tomita--Takesaki theory via probability theory. One can therefore envisage rewriting Tomita--Takesaki theory completely in probabilistic terms. Possible topics include a weight-version of Tomita--Takesaki theory, a reformulation of relative modular theory, and a stochastic-process reformulation of the paper \cite{DerezinskiJaksicPillet001}. In particular, a functional-integral version of the Golden--Thompson and Peierls--Bogoliubov inequalities would have a certain significance.

One can further envisage a formulation of the operator-algebra/probability-theory correspondence for fermions. The preprint \cite{RyosukeSato001} is explicitly aware of the paper \cite{KleinLandau001} and cites it. The paper \cite{ArakiMoriya001} develops a comprehensive discussion of equilibrium states for the CAR algebra on a lattice, and there is room to examine what can be obtained by its probabilistic rewriting. As for fermionic functional integrals, there is the textbook \cite{FeldmanKnorrerTrubowitz001} using integrals over Grassmann numbers, and there is also the formulation in the textbook \cite{BratteliRobinson004} using Brownian motion. What formulation is best is itself worth examining.

\subsection{Electron Models}\label{electron-models}

Starting with models analogous to the infinite-volume Hubbard model \cite{FrohlichUeltschi001,DanielUeltschi001} and the BCS model \cite{YoheiKashima001}, research on electron systems has developed significantly as a subject in its own right. The author himself does not possess wide expertise in this area, but would like to deepen the discussion of electron systems further.

\subsection{Quantum Measurement Theory}\label{quantum-measurement-theory}

In the paper \cite{GeoffreySewwell001} it is pointed out that the emergence of macroscopic states (phase cells) in quantum measurement has a structure analogous to that of phases and phase transitions in statistical mechanics. At least the BEC phase transition is understood as the emergence, in a state-dependent way, of the center of a von Neumann algebra, while the determination of the classical outcome of a measurement may likewise admit a description as an analogous decomposition structure. The precise formulation of this correspondence is not clear at present, but an attempt to understand the two within a unified operator-algebraic framework would be an interesting problem for the future.

\subsection{Formalization via Lean}\label{formalization-via-lean}

Recently a preprint on the formalization of relativistic quantum field theory \cite{DouglasHobackMeiNissim001} has appeared. Despite the difference between the relativistic and nonrelativistic settings, insofar as we argue primarily by means of functional integrals, there are both motivation and benefit in contributing to this project.

\bibliography{myref.bib}

\end{document}